\documentclass[aps,prl,reprint,superscriptaddress, longbibliography]{revtex4-2}

\usepackage[alsoload=hep]{siunitx}
\usepackage{amsmath}
\usepackage{amsbsy}
\usepackage{amssymb}
\usepackage{mathptmx, textcomp}
\usepackage{color}
\usepackage{braket}
\usepackage{graphicx}
\definecolor{bl}{rgb}{0, .1, .6}
\usepackage[colorlinks=true, citecolor = bl, linkcolor = bl, urlcolor=bl, pdfborder={0 0 0}]{hyperref}
\usepackage{notes2bib}

\DeclareSIUnit\gauss{G}

\begin{document}

\title{From superradiance to subradiance: exploring the many-body Dicke ladder}

\author{A. Glicenstein}
\email{antoine.glicenstein@institutoptique.fr}
\affiliation{Universit\'e Paris-Saclay, Institut d'Optique Graduate School, CNRS, 
Laboratoire Charles Fabry, 91127, Palaiseau, France}

\author{G. Ferioli}
\affiliation{Universit\'e Paris-Saclay, Institut d'Optique Graduate School, CNRS, 
Laboratoire Charles Fabry, 91127, Palaiseau, France}

\author{A. Browaeys}
\affiliation{Universit\'e Paris-Saclay, Institut d'Optique Graduate School, CNRS, 
Laboratoire Charles Fabry, 91127, Palaiseau, France}

\author{I. Ferrier-Barbut}
\affiliation{Universit\'e Paris-Saclay, Institut d'Optique Graduate School, CNRS, 
Laboratoire Charles Fabry, 91127, Palaiseau, France}

\begin{abstract}
We report a time-resolved study of collective emission in dense ensembles of two-level atoms. 
We compare, on the same sample, the build-up of superradiance and subradiance from the 
ensemble when driven by a strong laser. This allows us to measure the dynamics of the population 
of superradiant and subradiant states as a function of time. In particular we demonstrate the build up 
in time of subradiant states through the superradiant dynamics. This illustrates the dynamics of 
the many-body density matrix of superradiant ensembles of two-level atoms when departing from 
the ideal conditions of Dicke superradiance in which symmetry forbids the population of subradiant states.
\end{abstract}

\maketitle

Collective spontaneous emission of light by atomic ensembles 
has seen a renewed interest lately in the context of quantum technologies.  
Superradiance, i.e.~the emission of light at a rate enhanced with respect to the single atom case, 
was observed in many experiments \cite{skribanowitz1973observation,gross1976observation}. 
Recently, it was studied with cold atoms \cite{wang2007superradiance,paradis2008observation}, both in dilute clouds 
\cite{araujo2016superradiance,Roof2016observation,das2020subradiance} 
and dense ensembles \cite{ferioli2021laser}, also with atoms in cavity \cite{norcia2016superradiance} or trapped near 
waveguides and fibers \cite{Goban2015Superradiance,solano2017super,pennetta2021collective}. 
Its counterpart, subradiance, associated to a slow collective emission, 
has proven more challenging to observe \cite{devoe1996observation,mcguyer2015precise}. 
It has been demonstrated only in a handful of experiments using  dilute 
\cite{guerin2016subradiance, das2020subradiance, cipris2021subradiance} and dense 
\cite{ferioli2020storage,stiesdal2020observation} clouds of randomly distributed cold atoms. 
Here we address the question as to how superradiant and subradiant collective states 
are populated and evolve in laser-driven ensembles of two-level atoms. 

\begin{figure*}
\includegraphics[width=1\textwidth]{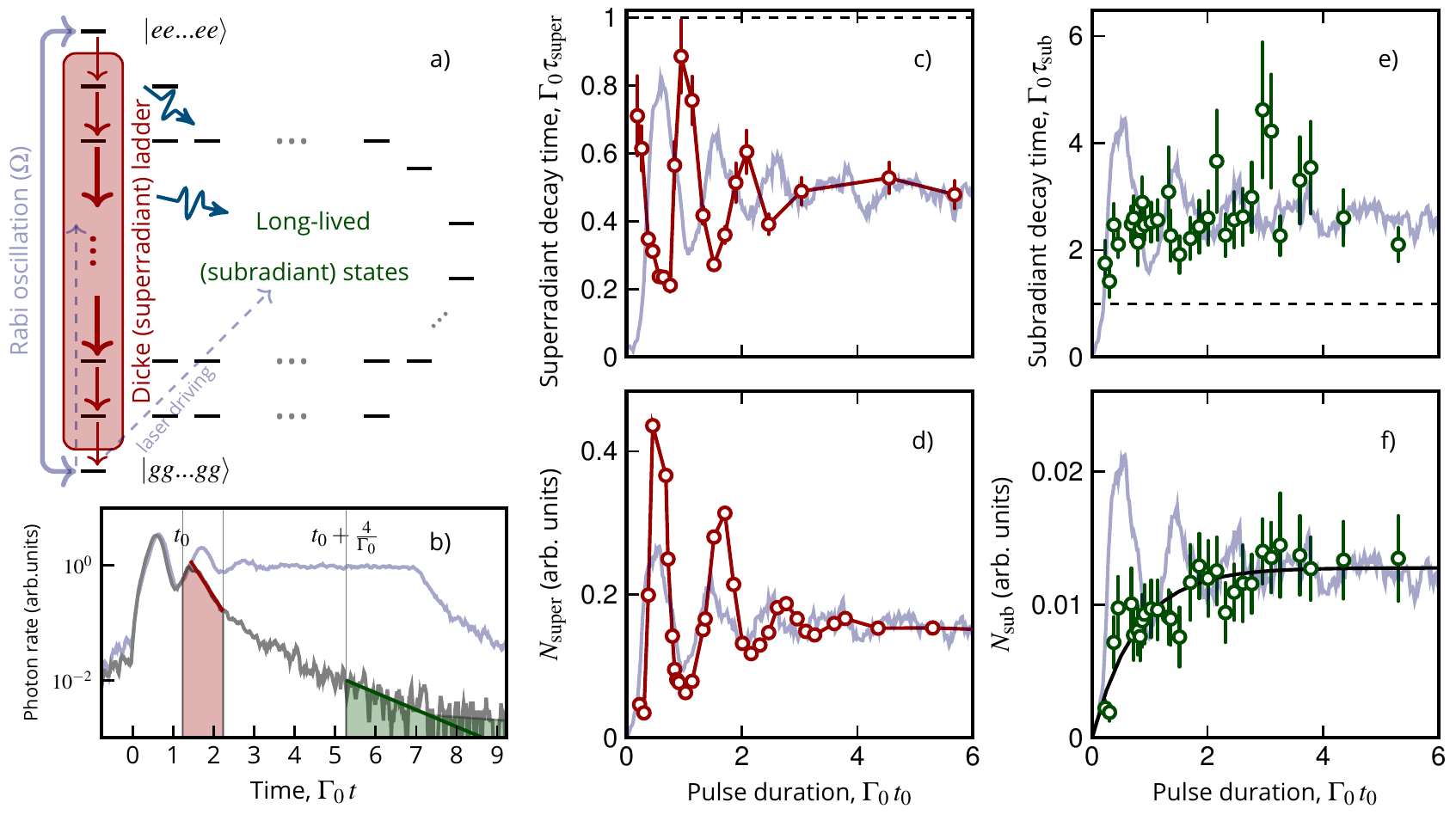}
\caption{{\bf Dynamics of super- and subradiant many-body states.} 
a) Representation of the ``Dicke ladder'' for our two-level ensembles. 
The superradiant states (highlighted in red) drive the cascade. 
The departure from ideal Dicke symmetries induces 
(i) a possibility for direct laser driving of the subradiant states (green) and 
(ii) a decay from superradiant states towards subradiant ones. 
c) and d) [resp.~e) and f)] show the dynamics under driving of the superradiant (resp.~subradiant) 
lifetime and photon count, 
extracted from the early (resp.~late) decay with timescales shown in b), 
see main text for details.} 
\label{fig1}
\end{figure*}

In Dicke's original proposal \cite{Dicke1954}, superradiance originates from an 
ensemble of $N$ two-level atoms (states $|g\rangle$ and $|e\rangle$),  placed in a volume small with respect 
to the transition wavelength and all initially inverted \cite{Gross1982}. In this scenario, only 
permutationally-symmetric superradiant states are populated during the decay.  
Contrarily, anti-symmetric subradiant states are {\it never} populated. 
Moreover, these subradiant states are strictly uncoupled even to an external drive, 
preventing the observation of subradiance. 
Experimental samples, however, depart from the symmetric conditions 
of the Dicke proposal, due to (i) the finite (or large) sample size
and (ii) the random positions of the atoms that leads to spatially varying dipole-dipole interactions
(the so-called van der Waals dephasing in~\cite{Gross1982}).
When sorted by the number $n$ of atoms in $|e\rangle$, the full set of states realizes
a ladder, with the resonant dipole-dipole interaction coupling the bare states inside a manifold corresponding to 
a given $n$ (see Fig.~\ref{fig1}a). 
Further, this breaking of symmetry makes it possible for symmetric states that, in the Dicke case, would lead only to a 
superradiant cascade to now decay towards subradiant ones. 
When all these $2^N$ states have to be considered, 
calculating the dynamics of the population of superradiant and subradiant states, 
following a population inversion or when driving the transition, is extremely 
challenging for large systems \cite{cipris2021subradiance,ferioli2021laser}.

To understand the conditions for the appearance of super- and subradiance and how the corresponding states are populated,  
we here perform experiments observing both phenomena on the \emph{same} sample. 
This is a dense, cigar-shaped cloud of two-level atoms 
with radial size smaller than the wavelength of their transition and axial size of a few wavelengths,
as in \cite{ferioli2020storage,ferioli2021laser}. 
By applying a global resonant excitation laser on the $g-e$ transition, 
and performing time-resolved fluorescence measurements during and after the excitation, we  
characterize both the superradiant and subradiant states in the same sample: 
we either nearly completely invert the system, or resolve the full dynamics during a long excitation pulse. 
We show that direct driving does populate subradiant states due to symmetry breaking,
and that a long excitation pulse populates 
slowly the subradiant manifold, in a process akin to optical pumping~\cite{cipris2021subradiance}, 
while superradiance is weakened.

Our setup, described in \cite{Glicenstein2021, Glicenstein2020}, operates an optical tweezers 
containing $N\simeq 4500$ $^{87}$Rb at a temperature of $\SI{650}{\micro\kelvin}$.  
The cloud's dimensions are $\sigma_{\text{ax}} = 15\lambda_0$ 
and $ \sigma_{\text{rad}} = 0.5\lambda_0$, 
where $\lambda_0 =780.2$~nm is the wavelength of the D2 line 
($\Gamma_0\simeq2\pi\,\times 6$~MHz, $I_{\rm sat}=\SI{1.67}{\milli\watt/\centi\meter^2}$). 
We isolate two Zeeman sub-levels with a $\SI{50}{\gauss}$ 
magnetic field perpendicularly to the cloud's axis \cite{ferioli2021laser}. 
The atoms are optically pumped in $|g\rangle=|5S_{1/2},F=2,m_F=-2 \rangle$ 
and are then excited perpendicularly to the cloud's axis (along the direction of the magnetic field), after being released from the trap, 
\footnote{The largest duration of the pulse we used is $\SI{150}{\nano\second}$ 
and the typical thermal velocity is $v_{th}\simeq\SI{0.25}{\meter/\second}$. 
The thermal expansion during the driving is $\simeq0.05\lambda_0$ and thus 
the atoms can be considered as frozen.},  
by a $\sigma_{-}$ polarized light resonant with $|e\rangle=| 5P_{3/2},F'=3,m_F'=-3\rangle$. 
The excitation beam size is  much larger than the atomic cloud.
The  intensity is $s=I/I_{\text{sat}}\simeq 75$ 
(Rabi frequency $\Omega\simeq6.5\Gamma_0$), with $\sim 10$\% shot-to-shot fluctuations. 
The temporal shape of the excitation is  controlled with a 
fibered electro-optical modulator with rising and falling times 
shorter than 1~ns, combined with two acousto-optical modulators 
to suppress residual light. 
The fluorescence emitted by the atoms along the cloud's axis is fiber-coupled to 
an avalanche photodiode operating in the single photon counting mode.

In order to investigate the influence of the initial atomic states on the collective decay, 
we tune the duration of the driving pulse, 
exploiting the Rabi oscillations  induced by the driving. 
If the pulse duration is small with respect to the spontaneous emission decay 
(occuring on a timescale $1/\Gamma_0$) the system is approximatively 
prepared in a coherent superposition 
$(\cos(\Omega t)\ket{g_n}+ie^{i{\bf k}_{\text{las}}\cdot {\bf R}_n}\sin(\Omega t)\ket{e_n})^{\otimes_n}$. 
In the Dicke limit, ${\bf k}_{\text{las}}\cdot {\bf R}_n\approx0$, and this initial state is coupled only to superradiant 
states since it is invariant under the exchange of two atoms. 
Conversely, for pulse durations long with respect to the decay and $\Omega\gg \Gamma_0$, the population reaches 
steady-state and the resulting density matrix is 
$\hat{\rho}_{\text{in}}= (\ket{e_n}\bra{e_n}+\ket{g_n}\bra{g_n})^{\otimes_n}/2^N$ 
in which super and subradiant states are equally populated \cite{cipris2021subradiance,Santos2021}. 
As we have recently shown~\cite{ferioli2021laser}, the axial emission during Rabi oscillations 
is enhanced by superradiance. However the evolution of the population of the excited state 
for the large Rabi frequency used here is only weakly modified and follows well the single atom behavior 
described by the optical Bloch equations (OBEs). 
By collecting the fluorescence emitted perpendicularly to the cloud's axis (not enhanced by superradiance) 
we are able to measure the evolution of the excited state fraction $n_e$ \cite{ferioli2021laser}. 
This evolution is also recovered when measuring the axial fluorescence at very low 
atom number where superradiance does not take place. 

To investigate the super and subradiant decay of our sample, 
we record the photon counts during a $\SI{1}{\nano\second}$ time bin, detected after having switched the excitation light off (at $t_0$) 
and separate the analysis in different time windows [seeFig.\,\ref{fig1}\,b)]. At short time, superradiance dominates  
and we extract the associated decay time $\tau_{\text{super}}$ by an exponential fit 
in the first $1/\Gamma_0\simeq\SI{26}{\nano\second}$ . 
We exclude the first $\SI{5}{\nano\second}$ from the fit to start it after the superradiant pulse occurring shortly after 
$t_0$, when close to the total population inversion \cite{ferioli2021laser}. 
We also record 
the total number of photons $N_{\rm super}$ during the first $\SI{26}{\nano\second}$. This count provides an indication of the population stored in superradiant states, which emit faster than the single atom lifetime. 
The results for  $\tau_{\text{super}}$ and $N_{\rm super}$ 
are reported in Fig.\,\ref{fig1}c),d) as a function of the duration of the excitation pulse. 
Both quantities oscillate at the same frequency, given by the Rabi frequency $\Omega=\sqrt{s/2}$. 
 The excited state fraction $n_e$, measured for low atom number, is reported as a reference in each panel of Fig.\,\ref{fig1}. 
The population of superradiant states oscillates in phase with respect to $n_e(t)$, 
as expected, while $\tau_{\text{super}}$ oscillates in quadrature 
(i.e.~the emission rate $\Gamma_{\text {super}} = 1/\tau_{\text{super}}$ oscillates in phase as expected). 

\begin{figure}[t]
\includegraphics[scale=1]{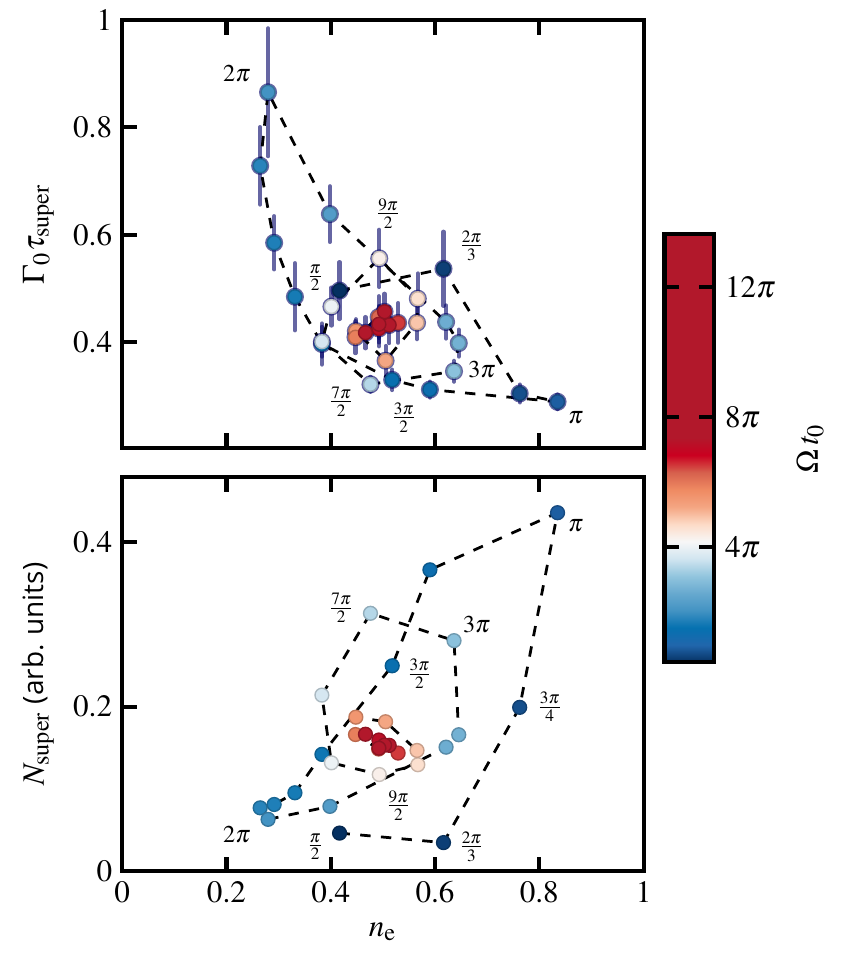}
\caption{Superradiant decay time (top), and photon count (bottom) 
 plotted as a function 
of excited state fraction $n_e$. The coloring indicates the time of the end of the pulse.
} 
\label{FigCorr}
\end{figure}

This observation is a confirmation that population inversion is the main drive for Dicke superradiance. 
But to deepen the analysis, we plot in Fig.\,\ref{FigCorr} the measured superradiant decay time 
and photon count, now as a function of the measured excited state population $n_e$. 
Strikingly, the data do not collapse on a single curve: for a given excited state fraction $n_e$, 
very different decay times and photon counts are observed. This stems from the fact  that superradiance 
arises not only from population inversion, but also from the emergence of in-phase correlations 
between atoms, of the form $e^{i{\bf k}_{\rm ax}.({\bf R}_m-{\bf R}_n)}\langle \hat\sigma_m^+\hat\sigma_n^-\rangle(t)$ 
where and $\hat\sigma^{+(-)}_n$ is the raising (lowering) operator for the state of atom $n$ 
with position ${\bf R}_n$, and ${\bf k}_{\rm ax}$ is the wavevector of the light emitted along the cloud axis. 
The superradiant states in our system are the ones for which these correlations are sizeable \cite{ferioli2021laser}. 
When the population is inverted, these correlations first grow in time before decaying 
via a superradiant cascade \cite{ferioli2021laser}. 
Thus depending at which stage of the collective Rabi cycles one measures the decay, 
a different dynamics unfolds. When we stop at $t_0$ the driving early in a Rabi cycle (say $\pi/2$ time for which $n_e=1/2$), 
the correlations have not had time to grow before $t_0$ and we observe a low photoemission rate and relatively slow final decay for $t\geqslant t_0$. On the other hand when we stop the driving at the end of a 
Rabi flop ($3\pi/2$, again with $n_e=1/2$) these correlations have had time to grow before $t_0$, 
so that a higher photoemission and faster decay is observed after $t_0$. 
This explains why at the same excited state fraction ($n_e=1/2$), 
different decay times and superradiant population can be observed. 
 

\begin{figure*}[htbp]
\includegraphics[width=\textwidth]{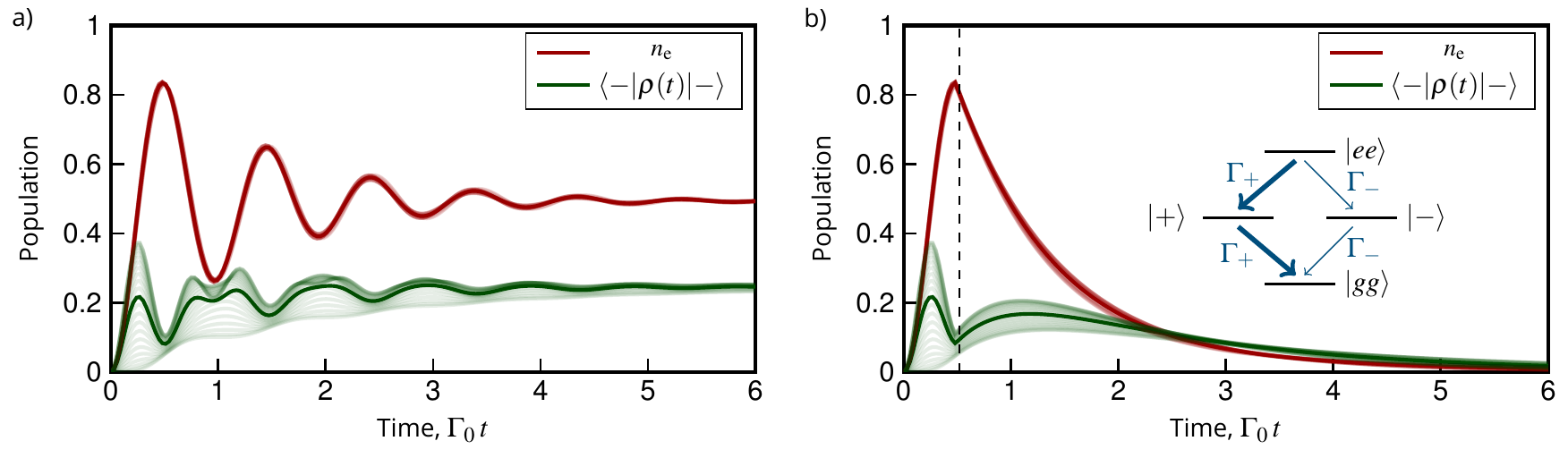}
\caption{\textbf{Simulation of the driving and decay of two atoms.} 
a) Excited state fraction (red) and population (green) of the subradiant state $\ket{-}$ 
during driving, for many different interatomic axis orientations (thin lines), and average (solid line). 
b) Same, but with driving switched-off at the $\pi$-pulse time. This allows to see initial direct driving 
of the population in $\ket{-}$ and then an increase in population due to decay from the doubly 
excited state $\ket{ee}$. The four states of the 2-atom system are shown as an inset in b).} 
\label{FigSimu}
\end{figure*}

Having probed the dynamics of superradiant states, we now move to the study of the subradiant decay occurring at later times after switching off the drive. 
For the same experimental data as above, we measure the decay time $\tau_{\rm sub}$
using an exponential fit for $t \geqslant t_0+4/\Gamma_0$ [see Fig.\,\ref{fig1}\,b)]. 
We now sum the photons recorded  
\emph{after} $t_0+4/\Gamma_0$. This count $N_{\rm sub}$ is related to the population of the long-lived states.
The results are reported in Fig.\,\ref{fig1}e),f). 
This decay time $\tau_{\rm sub}$ is nearly independent of the duration of the driving as the late decay mostly involve
subradiant states with similar decay times \cite{ferioli2020storage}. 
Next, contrarily to superradiance, 
we observe that the number of photons in the subradiant tail increases
nearly monotonically with time during the Rabi oscillations, saturating when the system 
reaches steady state. Importantly,  after a single Rabi cycle 
(at $t = \SI{25}{\nano\second}$), the population of the subradiant states has already grown.
This shows that, going through one cycle of Rabi oscillations 
(including through superradiant states) does lead to the population of subradiant states, due both to the decay 
towards subradiant states \cite{masson2020} and a direct driving from the ground state. 
This reveals the existence of a finite coupling between super 
and subradiant states due to the fact that, in our regime, 
the symmetry hypotheses on which Dicke's model is based are not satisfied. 
However, the subradiant population has \emph{not} reached its maximum value in a single cycle. 
Several such cycles are necessary to reach the maximum 
population of subradiant states. These measurements thus amount to a 
time-resolved observation of an effective optical pumping mechanism towards 
subradiant states through cycles in superradiant ones as suggested 
in \cite{cipris2021subradiance} for dilute clouds.

To gain intuition on the observations reported above, we have solved the master equation 
\cite{asenjo2017exponential} for $N=2$ interacting dipoles placed at an interatomic 
distance $\lambda/3$, that is the mean nearest neighbour distance in our experiment. 
The dipole-dipole interaction in the numerics is thus similar to the one of the experiment. 
For different times during the driving, we evaluate the excited state fraction
$n_e$, and that of the subradiant state 
$\bra{-}\rho(t)\ket{-}$ $\left[\ket{\pm}=\frac{1}{\sqrt{2}}(\ket{eg}\pm\ket{ge})\right]$. 
To partially  account for the randomness of the atomic spatial distribution, 
the simulations averaged over 1000 realizations, each obtained by changing 
the angle between the interatomic axis and the quantization axis, at a fixed interatomic distance. 

We report in Fig.\,\ref{FigSimu}a) the results of the simulation for an excitation pulse of 150~ns.  
They reproduce remarkably well the qualitative behavior observed in the experiment: 
while the excited state fraction $n_e$ oscillates, the population of the subradiant state is built through a different dynamics that involve both direct excitation and decay from the doubly excited state $\ket{ee}$.  
To highlight this, we report in Fig.\,\ref{FigSimu}b) the results of the simulations 
with a $\pi$-pulse where the population of $\ket{ee}$ is maximized: after the end of the driving, 
the population of subradiant state increases indicating that the symmetric state $\ket{ee}$ decays in part towards $\ket{-}$.
Depending on the relative orientation between the laser propagation axis and the interatomic axis, a direct excitation of $\ket{-}$ 
from the ground state is also possible, leading to a different dynamics for different orientations.
Our observations above indicate that both mechanisms take place in our many-body experiments. 

To conclude, we have explored the dynamics of super and subradiant modes hosted in 
the same atomic cloud of two-levels emitters close to the Dicke limit. 
By changing the duration of the excitation pulse, we prepare the system in different many-body density matrices, 
studying how super and subradiant states are populated during the driving and 
their interplay during the decay. We find that although superradiance 
is mainly governed by the excited state fraction, different correlations, that grow during
the drive lead to very different superradiant decays. 
We resolved in time how subradiant modes are populated via both a direct coupling and 
an optical-pumping-like mechanism through the Dicke ladder. 
Elucidating the correlation growth dynamics will be key to exploring the phase diagram of ordered superradiant ensembles \cite{olmos2014steady,parmee2018phases,parmee2020signatures,lewis2021characterizing}.
Furthermore, future experimental studies should
aim at measuring the predicted growth of entanglement associated to subradiance, e.g. 
by measuring intensity correlation of the fluorescence light \cite{Santos2021}.

\begin{acknowledgements}
We thank F. Robicheaux and R.T. Sutherland for discussions.
This project has received funding from the European Union's Horizon 2020 
research and innovation program under Grant Agreement No. 817482 (PASQuanS) 
and by the Region Ile-de-France in the framework of DIM SIRTEQ (project DSHAPE). 
A. G. is supported by the Del\'egation G\'en\'erale de l'Armement Fellowship No. 2018.60.0027.
\end{acknowledgements}

\bibliography{subradiance2}

\begin{thebibliography}{30}%
\makeatletter
\providecommand \@ifxundefined [1]{%
 \@ifx{#1\undefined}
}%
\providecommand \@ifnum [1]{%
 \ifnum #1\expandafter \@firstoftwo
 \else \expandafter \@secondoftwo
 \fi
}%
\providecommand \@ifx [1]{%
 \ifx #1\expandafter \@firstoftwo
 \else \expandafter \@secondoftwo
 \fi
}%
\providecommand \natexlab [1]{#1}%
\providecommand \enquote  [1]{``#1''}%
\providecommand \bibnamefont  [1]{#1}%
\providecommand \bibfnamefont [1]{#1}%
\providecommand \citenamefont [1]{#1}%
\providecommand \href@noop [0]{\@secondoftwo}%
\providecommand \href [0]{\begingroup \@sanitize@url \@href}%
\providecommand \@href[1]{\@@startlink{#1}\@@href}%
\providecommand \@@href[1]{\endgroup#1\@@endlink}%
\providecommand \@sanitize@url [0]{\catcode `\\12\catcode `\$12\catcode
  `\&12\catcode `\#12\catcode `\^12\catcode `\_12\catcode `\%12\relax}%
\providecommand \@@startlink[1]{}%
\providecommand \@@endlink[0]{}%
\providecommand \url  [0]{\begingroup\@sanitize@url \@url }%
\providecommand \@url [1]{\endgroup\@href {#1}{\urlprefix }}%
\providecommand \urlprefix  [0]{URL }%
\providecommand \Eprint [0]{\href }%
\providecommand \doibase [0]{https://doi.org/}%
\providecommand \selectlanguage [0]{\@gobble}%
\providecommand \bibinfo  [0]{\@secondoftwo}%
\providecommand \bibfield  [0]{\@secondoftwo}%
\providecommand \translation [1]{[#1]}%
\providecommand \BibitemOpen [0]{}%
\providecommand \bibitemStop [0]{}%
\providecommand \bibitemNoStop [0]{.\EOS\space}%
\providecommand \EOS [0]{\spacefactor3000\relax}%
\providecommand \BibitemShut  [1]{\csname bibitem#1\endcsname}%
\let\auto@bib@innerbib\@empty
\bibitem [{\citenamefont {Skribanowitz}\ \emph {et~al.}(1973)\citenamefont
  {Skribanowitz}, \citenamefont {Herman}, \citenamefont {MacGillivray},\ and\
  \citenamefont {Feld}}]{skribanowitz1973observation}%
  \BibitemOpen
  \bibfield  {author} {\bibinfo {author} {\bibfnamefont {N.}~\bibnamefont
  {Skribanowitz}}, \bibinfo {author} {\bibfnamefont {I.~P.}\ \bibnamefont
  {Herman}}, \bibinfo {author} {\bibfnamefont {J.~C.}\ \bibnamefont
  {MacGillivray}},\ and\ \bibinfo {author} {\bibfnamefont {M.~S.}\ \bibnamefont
  {Feld}},\ }\bibfield  {title} {\bibinfo {title} {Observation of dicke
  superradiance in optically pumped hf gas},\ }\href
  {https://doi.org/10.1103/PhysRevLett.30.309} {\bibfield  {journal} {\bibinfo
  {journal} {Phys. Rev. Lett.}\ }\textbf {\bibinfo {volume} {30}},\ \bibinfo
  {pages} {309} (\bibinfo {year} {1973})}\BibitemShut {NoStop}%
\bibitem [{\citenamefont {Gross}\ \emph {et~al.}(1976)\citenamefont {Gross},
  \citenamefont {Fabre}, \citenamefont {Pillet},\ and\ \citenamefont
  {Haroche}}]{gross1976observation}%
  \BibitemOpen
  \bibfield  {author} {\bibinfo {author} {\bibfnamefont {M.}~\bibnamefont
  {Gross}}, \bibinfo {author} {\bibfnamefont {C.}~\bibnamefont {Fabre}},
  \bibinfo {author} {\bibfnamefont {P.}~\bibnamefont {Pillet}},\ and\ \bibinfo
  {author} {\bibfnamefont {S.}~\bibnamefont {Haroche}},\ }\bibfield  {title}
  {\bibinfo {title} {Observation of near-infrared dicke superradiance on
  cascading transitions in atomic sodium},\ }\href
  {https://doi.org/10.1103/PhysRevLett.36.1035} {\bibfield  {journal} {\bibinfo
   {journal} {Phys. Rev. Lett.}\ }\textbf {\bibinfo {volume} {36}},\ \bibinfo
  {pages} {1035} (\bibinfo {year} {1976})}\BibitemShut {NoStop}%
\bibitem [{\citenamefont {Wang}\ \emph {et~al.}(2007)\citenamefont {Wang},
  \citenamefont {Yelin}, \citenamefont {C\^ot\'e}, \citenamefont {Eyler},
  \citenamefont {Farooqi}, \citenamefont {Gould}, \citenamefont
  {Ko\ifmmode~\check{s}\else \v{s}\fi{}trun}, \citenamefont {Tong},\ and\
  \citenamefont {Vrinceanu}}]{wang2007superradiance}%
  \BibitemOpen
  \bibfield  {author} {\bibinfo {author} {\bibfnamefont {T.}~\bibnamefont
  {Wang}}, \bibinfo {author} {\bibfnamefont {S.~F.}\ \bibnamefont {Yelin}},
  \bibinfo {author} {\bibfnamefont {R.}~\bibnamefont {C\^ot\'e}}, \bibinfo
  {author} {\bibfnamefont {E.~E.}\ \bibnamefont {Eyler}}, \bibinfo {author}
  {\bibfnamefont {S.~M.}\ \bibnamefont {Farooqi}}, \bibinfo {author}
  {\bibfnamefont {P.~L.}\ \bibnamefont {Gould}}, \bibinfo {author}
  {\bibfnamefont {M.}~\bibnamefont {Ko\ifmmode~\check{s}\else \v{s}\fi{}trun}},
  \bibinfo {author} {\bibfnamefont {D.}~\bibnamefont {Tong}},\ and\ \bibinfo
  {author} {\bibfnamefont {D.}~\bibnamefont {Vrinceanu}},\ }\bibfield  {title}
  {\bibinfo {title} {Superradiance in ultracold rydberg gases},\ }\href
  {https://doi.org/10.1103/PhysRevA.75.033802} {\bibfield  {journal} {\bibinfo
  {journal} {Phys. Rev. A}\ }\textbf {\bibinfo {volume} {75}},\ \bibinfo
  {pages} {033802} (\bibinfo {year} {2007})}\BibitemShut {NoStop}%
\bibitem [{\citenamefont {Paradis}\ \emph {et~al.}(2008)\citenamefont
  {Paradis}, \citenamefont {Barrett}, \citenamefont {Kumarakrishnan},
  \citenamefont {Zhang},\ and\ \citenamefont
  {Raithel}}]{paradis2008observation}%
  \BibitemOpen
  \bibfield  {author} {\bibinfo {author} {\bibfnamefont {E.}~\bibnamefont
  {Paradis}}, \bibinfo {author} {\bibfnamefont {B.}~\bibnamefont {Barrett}},
  \bibinfo {author} {\bibfnamefont {A.}~\bibnamefont {Kumarakrishnan}},
  \bibinfo {author} {\bibfnamefont {R.}~\bibnamefont {Zhang}},\ and\ \bibinfo
  {author} {\bibfnamefont {G.}~\bibnamefont {Raithel}},\ }\bibfield  {title}
  {\bibinfo {title} {Observation of superfluorescent emissions from
  laser-cooled atoms},\ }\href {https://doi.org/10.1103/PhysRevA.77.043419}
  {\bibfield  {journal} {\bibinfo  {journal} {Phys. Rev. A}\ }\textbf {\bibinfo
  {volume} {77}},\ \bibinfo {pages} {043419} (\bibinfo {year}
  {2008})}\BibitemShut {NoStop}%
\bibitem [{\citenamefont {Ara\'ujo}\ \emph {et~al.}(2016)\citenamefont
  {Ara\'ujo}, \citenamefont {Kre\ifmmode \check{s}\else
  \v{s}\fi{}i\ifmmode~\acute{c}\else \'{c}\fi{}}, \citenamefont {Kaiser},\ and\
  \citenamefont {Guerin}}]{araujo2016superradiance}%
  \BibitemOpen
  \bibfield  {author} {\bibinfo {author} {\bibfnamefont {M.~O.}\ \bibnamefont
  {Ara\'ujo}}, \bibinfo {author} {\bibfnamefont {I.}~\bibnamefont {Kre\ifmmode
  \check{s}\else \v{s}\fi{}i\ifmmode~\acute{c}\else \'{c}\fi{}}}, \bibinfo
  {author} {\bibfnamefont {R.}~\bibnamefont {Kaiser}},\ and\ \bibinfo {author}
  {\bibfnamefont {W.}~\bibnamefont {Guerin}},\ }\bibfield  {title} {\bibinfo
  {title} {Superradiance in a large and dilute cloud of cold atoms in the
  linear-optics regime},\ }\href
  {https://doi.org/10.1103/PhysRevLett.117.073002} {\bibfield  {journal}
  {\bibinfo  {journal} {Phys. Rev. Lett.}\ }\textbf {\bibinfo {volume} {117}},\
  \bibinfo {pages} {073002} (\bibinfo {year} {2016})}\BibitemShut {NoStop}%
\bibitem [{\citenamefont {Roof}\ \emph {et~al.}(2016)\citenamefont {Roof},
  \citenamefont {Kemp}, \citenamefont {Havey},\ and\ \citenamefont
  {Sokolov}}]{Roof2016observation}%
  \BibitemOpen
  \bibfield  {author} {\bibinfo {author} {\bibfnamefont {S.~J.}\ \bibnamefont
  {Roof}}, \bibinfo {author} {\bibfnamefont {K.~J.}\ \bibnamefont {Kemp}},
  \bibinfo {author} {\bibfnamefont {M.~D.}\ \bibnamefont {Havey}},\ and\
  \bibinfo {author} {\bibfnamefont {I.~M.}\ \bibnamefont {Sokolov}},\
  }\bibfield  {title} {\bibinfo {title} {Observation of single-photon
  superradiance and the cooperative lamb shift in an extended sample of cold
  atoms},\ }\href {https://doi.org/10.1103/PhysRevLett.117.073003} {\bibfield
  {journal} {\bibinfo  {journal} {Phys. Rev. Lett.}\ }\textbf {\bibinfo
  {volume} {117}},\ \bibinfo {pages} {073003} (\bibinfo {year}
  {2016})}\BibitemShut {NoStop}%
\bibitem [{\citenamefont {Das}\ \emph {et~al.}(2020)\citenamefont {Das},
  \citenamefont {Lemberger},\ and\ \citenamefont {Yavuz}}]{das2020subradiance}%
  \BibitemOpen
  \bibfield  {author} {\bibinfo {author} {\bibfnamefont {D.}~\bibnamefont
  {Das}}, \bibinfo {author} {\bibfnamefont {B.}~\bibnamefont {Lemberger}},\
  and\ \bibinfo {author} {\bibfnamefont {D.~D.}\ \bibnamefont {Yavuz}},\
  }\bibfield  {title} {\bibinfo {title} {Subradiance and
  superradiance-to-subradiance transition in dilute atomic clouds},\ }\href
  {https://doi.org/10.1103/PhysRevA.102.043708} {\bibfield  {journal} {\bibinfo
   {journal} {Phys. Rev. A}\ }\textbf {\bibinfo {volume} {102}},\ \bibinfo
  {pages} {043708} (\bibinfo {year} {2020})}\BibitemShut {NoStop}%
\bibitem [{\citenamefont {Ferioli}\ \emph
  {et~al.}(2021{\natexlab{a}})\citenamefont {Ferioli}, \citenamefont
  {Glicenstein}, \citenamefont {Robicheaux}, \citenamefont {Sutherland},
  \citenamefont {Browaeys},\ and\ \citenamefont
  {Ferrier-Barbut}}]{ferioli2021laser}%
  \BibitemOpen
  \bibfield  {author} {\bibinfo {author} {\bibfnamefont {G.}~\bibnamefont
  {Ferioli}}, \bibinfo {author} {\bibfnamefont {A.}~\bibnamefont
  {Glicenstein}}, \bibinfo {author} {\bibfnamefont {F.}~\bibnamefont
  {Robicheaux}}, \bibinfo {author} {\bibfnamefont {R.~T.}\ \bibnamefont
  {Sutherland}}, \bibinfo {author} {\bibfnamefont {A.}~\bibnamefont
  {Browaeys}},\ and\ \bibinfo {author} {\bibfnamefont {I.}~\bibnamefont
  {Ferrier-Barbut}},\ }\bibfield  {title} {\bibinfo {title} {Laser-driven
  superradiant ensembles of two-level atoms near dicke regime},\ }\href
  {https://doi.org/10.1103/PhysRevLett.127.243602} {\bibfield  {journal}
  {\bibinfo  {journal} {Phys. Rev. Lett.}\ }\textbf {\bibinfo {volume} {127}},\
  \bibinfo {pages} {243602} (\bibinfo {year} {2021}{\natexlab{a}})}\BibitemShut
  {NoStop}%
\bibitem [{\citenamefont {Norcia}\ \emph {et~al.}(2016)\citenamefont {Norcia},
  \citenamefont {Winchester}, \citenamefont {Cline},\ and\ \citenamefont
  {Thompson}}]{norcia2016superradiance}%
  \BibitemOpen
  \bibfield  {author} {\bibinfo {author} {\bibfnamefont {M.~A.}\ \bibnamefont
  {Norcia}}, \bibinfo {author} {\bibfnamefont {M.~N.}\ \bibnamefont
  {Winchester}}, \bibinfo {author} {\bibfnamefont {J.~R.~K.}\ \bibnamefont
  {Cline}},\ and\ \bibinfo {author} {\bibfnamefont {J.~K.}\ \bibnamefont
  {Thompson}},\ }\bibfield  {title} {\bibinfo {title} {Superradiance on the
  millihertz linewidth strontium clock transition},\ }\bibfield  {journal}
  {\bibinfo  {journal} {Science Advances}\ }\textbf {\bibinfo {volume} {2}},\
  \href {https://doi.org/10.1126/sciadv.1601231} {10.1126/sciadv.1601231}
  (\bibinfo {year} {2016}),\ \Eprint
  {https://arxiv.org/abs/https://advances.sciencemag.org/content/2/10/e1601231.full.pdf}
  {https://advances.sciencemag.org/content/2/10/e1601231.full.pdf} \BibitemShut
  {NoStop}%
\bibitem [{\citenamefont {Goban}\ \emph {et~al.}(2015)\citenamefont {Goban},
  \citenamefont {Hung}, \citenamefont {Hood}, \citenamefont {Yu}, \citenamefont
  {Muniz}, \citenamefont {Painter},\ and\ \citenamefont
  {Kimble}}]{Goban2015Superradiance}%
  \BibitemOpen
  \bibfield  {author} {\bibinfo {author} {\bibfnamefont {A.}~\bibnamefont
  {Goban}}, \bibinfo {author} {\bibfnamefont {C.-L.}\ \bibnamefont {Hung}},
  \bibinfo {author} {\bibfnamefont {J.~D.}\ \bibnamefont {Hood}}, \bibinfo
  {author} {\bibfnamefont {S.-P.}\ \bibnamefont {Yu}}, \bibinfo {author}
  {\bibfnamefont {J.~A.}\ \bibnamefont {Muniz}}, \bibinfo {author}
  {\bibfnamefont {O.}~\bibnamefont {Painter}},\ and\ \bibinfo {author}
  {\bibfnamefont {H.~J.}\ \bibnamefont {Kimble}},\ }\bibfield  {title}
  {\bibinfo {title} {Superradiance for atoms trapped along a photonic crystal
  waveguide},\ }\href {https://doi.org/10.1103/PhysRevLett.115.063601}
  {\bibfield  {journal} {\bibinfo  {journal} {Phys. Rev. Lett.}\ }\textbf
  {\bibinfo {volume} {115}},\ \bibinfo {pages} {063601} (\bibinfo {year}
  {2015})}\BibitemShut {NoStop}%
\bibitem [{\citenamefont {Solano}\ \emph {et~al.}(2017)\citenamefont {Solano},
  \citenamefont {Barberis-Blostein}, \citenamefont {Fatemi}, \citenamefont
  {Orozco},\ and\ \citenamefont {Rolston}}]{solano2017super}%
  \BibitemOpen
  \bibfield  {author} {\bibinfo {author} {\bibfnamefont {P.}~\bibnamefont
  {Solano}}, \bibinfo {author} {\bibfnamefont {P.}~\bibnamefont
  {Barberis-Blostein}}, \bibinfo {author} {\bibfnamefont {F.~K.}\ \bibnamefont
  {Fatemi}}, \bibinfo {author} {\bibfnamefont {L.~A.}\ \bibnamefont {Orozco}},\
  and\ \bibinfo {author} {\bibfnamefont {S.~L.}\ \bibnamefont {Rolston}},\
  }\bibfield  {title} {\bibinfo {title} {Super-radiance reveals infinite-range
  dipole interactions through a nanofiber},\ }\href
  {https://doi.org/10.1038/s41467-017-01994-3} {\bibfield  {journal} {\bibinfo
  {journal} {Nature Communications}\ }\textbf {\bibinfo {volume} {8}},\
  \bibinfo {pages} {1857} (\bibinfo {year} {2017})}\BibitemShut {NoStop}%
\bibitem [{\citenamefont {Pennetta}\ \emph {et~al.}(2021)\citenamefont
  {Pennetta}, \citenamefont {Blaha}, \citenamefont {Johnson}, \citenamefont
  {Lechner}, \citenamefont {Schneeweiss}, \citenamefont {Volz},\ and\
  \citenamefont {Rauschenbeutel}}]{pennetta2021collective}%
  \BibitemOpen
  \bibfield  {author} {\bibinfo {author} {\bibfnamefont {R.}~\bibnamefont
  {Pennetta}}, \bibinfo {author} {\bibfnamefont {M.}~\bibnamefont {Blaha}},
  \bibinfo {author} {\bibfnamefont {A.}~\bibnamefont {Johnson}}, \bibinfo
  {author} {\bibfnamefont {D.}~\bibnamefont {Lechner}}, \bibinfo {author}
  {\bibfnamefont {P.}~\bibnamefont {Schneeweiss}}, \bibinfo {author}
  {\bibfnamefont {J.}~\bibnamefont {Volz}},\ and\ \bibinfo {author}
  {\bibfnamefont {A.}~\bibnamefont {Rauschenbeutel}},\ }\href@noop {} {\bibinfo
  {title} {Collective radiative dynamics of an ensemble of cold atoms coupled
  to an optical waveguide}} (\bibinfo {year} {2021}),\ \Eprint
  {https://arxiv.org/abs/2109.00860} {arXiv:2109.00860 [quant-ph]} \BibitemShut
  {NoStop}%
\bibitem [{\citenamefont {DeVoe}\ and\ \citenamefont
  {Brewer}(1996)}]{devoe1996observation}%
  \BibitemOpen
  \bibfield  {author} {\bibinfo {author} {\bibfnamefont {R.~G.}\ \bibnamefont
  {DeVoe}}\ and\ \bibinfo {author} {\bibfnamefont {R.~G.}\ \bibnamefont
  {Brewer}},\ }\bibfield  {title} {\bibinfo {title} {Observation of
  superradiant and subradiant spontaneous emission of two trapped ions},\
  }\href {https://doi.org/10.1103/PhysRevLett.76.2049} {\bibfield  {journal}
  {\bibinfo  {journal} {Phys. Rev. Lett.}\ }\textbf {\bibinfo {volume} {76}},\
  \bibinfo {pages} {2049} (\bibinfo {year} {1996})}\BibitemShut {NoStop}%
\bibitem [{\citenamefont {McGuyer}\ \emph {et~al.}(2015)\citenamefont
  {McGuyer}, \citenamefont {McDonald}, \citenamefont {Iwata}, \citenamefont
  {Tarallo}, \citenamefont {Skomorowski}, \citenamefont {Moszynski},\ and\
  \citenamefont {Zelevinsky}}]{mcguyer2015precise}%
  \BibitemOpen
  \bibfield  {author} {\bibinfo {author} {\bibfnamefont {B.~H.}\ \bibnamefont
  {McGuyer}}, \bibinfo {author} {\bibfnamefont {M.}~\bibnamefont {McDonald}},
  \bibinfo {author} {\bibfnamefont {G.~Z.}\ \bibnamefont {Iwata}}, \bibinfo
  {author} {\bibfnamefont {M.~G.}\ \bibnamefont {Tarallo}}, \bibinfo {author}
  {\bibfnamefont {W.}~\bibnamefont {Skomorowski}}, \bibinfo {author}
  {\bibfnamefont {R.}~\bibnamefont {Moszynski}},\ and\ \bibinfo {author}
  {\bibfnamefont {T.}~\bibnamefont {Zelevinsky}},\ }\bibfield  {title}
  {\bibinfo {title} {Precise study of asymptotic physics with subradiant
  ultracold molecules},\ }\href {https://doi.org/10.1038/nphys3182} {\bibfield
  {journal} {\bibinfo  {journal} {Nature Physics}\ }\textbf {\bibinfo {volume}
  {11}},\ \bibinfo {pages} {32} (\bibinfo {year} {2015})}\BibitemShut {NoStop}%
\bibitem [{\citenamefont {Guerin}\ \emph {et~al.}(2016)\citenamefont {Guerin},
  \citenamefont {Ara\'ujo},\ and\ \citenamefont
  {Kaiser}}]{guerin2016subradiance}%
  \BibitemOpen
  \bibfield  {author} {\bibinfo {author} {\bibfnamefont {W.}~\bibnamefont
  {Guerin}}, \bibinfo {author} {\bibfnamefont {M.~O.}\ \bibnamefont
  {Ara\'ujo}},\ and\ \bibinfo {author} {\bibfnamefont {R.}~\bibnamefont
  {Kaiser}},\ }\bibfield  {title} {\bibinfo {title} {Subradiance in a large
  cloud of cold atoms},\ }\href
  {https://doi.org/10.1103/PhysRevLett.116.083601} {\bibfield  {journal}
  {\bibinfo  {journal} {Phys. Rev. Lett.}\ }\textbf {\bibinfo {volume} {116}},\
  \bibinfo {pages} {083601} (\bibinfo {year} {2016})}\BibitemShut {NoStop}%
\bibitem [{\citenamefont {Cipris}\ \emph {et~al.}(2021)\citenamefont {Cipris},
  \citenamefont {Moreira}, \citenamefont {do~Espirito~Santo}, \citenamefont
  {Weiss}, \citenamefont {Villas-Boas}, \citenamefont {Kaiser}, \citenamefont
  {Guerin},\ and\ \citenamefont {Bachelard}}]{cipris2021subradiance}%
  \BibitemOpen
  \bibfield  {author} {\bibinfo {author} {\bibfnamefont {A.}~\bibnamefont
  {Cipris}}, \bibinfo {author} {\bibfnamefont {N.~A.}\ \bibnamefont {Moreira}},
  \bibinfo {author} {\bibfnamefont {T.~S.}\ \bibnamefont {do~Espirito~Santo}},
  \bibinfo {author} {\bibfnamefont {P.}~\bibnamefont {Weiss}}, \bibinfo
  {author} {\bibfnamefont {C.~J.}\ \bibnamefont {Villas-Boas}}, \bibinfo
  {author} {\bibfnamefont {R.}~\bibnamefont {Kaiser}}, \bibinfo {author}
  {\bibfnamefont {W.}~\bibnamefont {Guerin}},\ and\ \bibinfo {author}
  {\bibfnamefont {R.}~\bibnamefont {Bachelard}},\ }\bibfield  {title} {\bibinfo
  {title} {Subradiance with saturated atoms: Population enhancement of the
  long-lived states},\ }\href {https://doi.org/10.1103/PhysRevLett.126.103604}
  {\bibfield  {journal} {\bibinfo  {journal} {Phys. Rev. Lett.}\ }\textbf
  {\bibinfo {volume} {126}},\ \bibinfo {pages} {103604} (\bibinfo {year}
  {2021})}\BibitemShut {NoStop}%
\bibitem [{\citenamefont {Ferioli}\ \emph
  {et~al.}(2021{\natexlab{b}})\citenamefont {Ferioli}, \citenamefont
  {Glicenstein}, \citenamefont {Henriet}, \citenamefont {Ferrier-Barbut},\ and\
  \citenamefont {Browaeys}}]{ferioli2020storage}%
  \BibitemOpen
  \bibfield  {author} {\bibinfo {author} {\bibfnamefont {G.}~\bibnamefont
  {Ferioli}}, \bibinfo {author} {\bibfnamefont {A.}~\bibnamefont
  {Glicenstein}}, \bibinfo {author} {\bibfnamefont {L.}~\bibnamefont
  {Henriet}}, \bibinfo {author} {\bibfnamefont {I.}~\bibnamefont
  {Ferrier-Barbut}},\ and\ \bibinfo {author} {\bibfnamefont {A.}~\bibnamefont
  {Browaeys}},\ }\bibfield  {title} {\bibinfo {title} {Storage and release of
  subradiant excitations in a dense atomic cloud},\ }\href
  {https://doi.org/10.1103/PhysRevX.11.021031} {\bibfield  {journal} {\bibinfo
  {journal} {Phys. Rev. X}\ }\textbf {\bibinfo {volume} {11}},\ \bibinfo
  {pages} {021031} (\bibinfo {year} {2021}{\natexlab{b}})}\BibitemShut
  {NoStop}%
\bibitem [{\citenamefont {Stiesdal}\ \emph {et~al.}(2020)\citenamefont
  {Stiesdal}, \citenamefont {Busche}, \citenamefont {Kumlin}, \citenamefont
  {Kleinbeck}, \citenamefont {B\"uchler},\ and\ \citenamefont
  {Hofferberth}}]{stiesdal2020observation}%
  \BibitemOpen
  \bibfield  {author} {\bibinfo {author} {\bibfnamefont {N.}~\bibnamefont
  {Stiesdal}}, \bibinfo {author} {\bibfnamefont {H.}~\bibnamefont {Busche}},
  \bibinfo {author} {\bibfnamefont {J.}~\bibnamefont {Kumlin}}, \bibinfo
  {author} {\bibfnamefont {K.}~\bibnamefont {Kleinbeck}}, \bibinfo {author}
  {\bibfnamefont {H.~P.}\ \bibnamefont {B\"uchler}},\ and\ \bibinfo {author}
  {\bibfnamefont {S.}~\bibnamefont {Hofferberth}},\ }\bibfield  {title}
  {\bibinfo {title} {Observation of collective decay dynamics of a single
  rydberg superatom},\ }\href
  {https://doi.org/10.1103/PhysRevResearch.2.043339} {\bibfield  {journal}
  {\bibinfo  {journal} {Phys. Rev. Research}\ }\textbf {\bibinfo {volume}
  {2}},\ \bibinfo {pages} {043339} (\bibinfo {year} {2020})}\BibitemShut
  {NoStop}%
\bibitem [{\citenamefont {Dicke}(1954)}]{Dicke1954}%
  \BibitemOpen
  \bibfield  {author} {\bibinfo {author} {\bibfnamefont {R.~H.}\ \bibnamefont
  {Dicke}},\ }\bibfield  {title} {\bibinfo {title} {Coherence in spontaneous
  radiation processes},\ }\href {https://doi.org/10.1103/PhysRev.93.99}
  {\bibfield  {journal} {\bibinfo  {journal} {Phys. Rev.}\ }\textbf {\bibinfo
  {volume} {93}},\ \bibinfo {pages} {99} (\bibinfo {year} {1954})}\BibitemShut
  {NoStop}%
\bibitem [{\citenamefont {Gross}\ and\ \citenamefont
  {Haroche}(1982)}]{Gross1982}%
  \BibitemOpen
  \bibfield  {author} {\bibinfo {author} {\bibfnamefont {M.}~\bibnamefont
  {Gross}}\ and\ \bibinfo {author} {\bibfnamefont {S.}~\bibnamefont
  {Haroche}},\ }\bibfield  {title} {\bibinfo {title} {Superradiance: An essay
  on the theory of collective spontaneous emission},\ }\href
  {https://doi.org/https://doi.org/10.1016/0370-1573(82)90102-8} {\bibfield
  {journal} {\bibinfo  {journal} {Physics Reports}\ }\textbf {\bibinfo {volume}
  {93}},\ \bibinfo {pages} {301 } (\bibinfo {year} {1982})}\BibitemShut
  {NoStop}%
\bibitem [{\citenamefont {Glicenstein}\ \emph {et~al.}(2021)\citenamefont
  {Glicenstein}, \citenamefont {Ferioli}, \citenamefont {Brossard},
  \citenamefont {Sortais}, \citenamefont {Barredo}, \citenamefont {Nogrette},
  \citenamefont {Ferrier-Barbut},\ and\ \citenamefont
  {Browaeys}}]{Glicenstein2021}%
  \BibitemOpen
  \bibfield  {author} {\bibinfo {author} {\bibfnamefont {A.}~\bibnamefont
  {Glicenstein}}, \bibinfo {author} {\bibfnamefont {G.}~\bibnamefont
  {Ferioli}}, \bibinfo {author} {\bibfnamefont {L.}~\bibnamefont {Brossard}},
  \bibinfo {author} {\bibfnamefont {Y.~R.~P.}\ \bibnamefont {Sortais}},
  \bibinfo {author} {\bibfnamefont {D.}~\bibnamefont {Barredo}}, \bibinfo
  {author} {\bibfnamefont {F.}~\bibnamefont {Nogrette}}, \bibinfo {author}
  {\bibfnamefont {I.}~\bibnamefont {Ferrier-Barbut}},\ and\ \bibinfo {author}
  {\bibfnamefont {A.}~\bibnamefont {Browaeys}},\ }\bibfield  {title} {\bibinfo
  {title} {{Preparation of one-dimensional chains and dense cold atomic clouds
  with a high numerical aperture four-lens system}},\ }\href
  {https://doi.org/10.1103/physreva.103.043301} {\bibfield  {journal} {\bibinfo
   {journal} {Physical Review A}\ }\textbf {\bibinfo {volume} {103}},\ \bibinfo
  {pages} {43301} (\bibinfo {year} {2021})}\BibitemShut {NoStop}%
\bibitem [{\citenamefont {Glicenstein}\ \emph {et~al.}(2020)\citenamefont
  {Glicenstein}, \citenamefont {Ferioli}, \citenamefont {\ifmmode
  \check{S}\else \v{S}\fi{}ibali\ifmmode~\acute{c}\else \'{c}\fi{}},
  \citenamefont {Brossard}, \citenamefont {Ferrier-Barbut},\ and\ \citenamefont
  {Browaeys}}]{Glicenstein2020}%
  \BibitemOpen
  \bibfield  {author} {\bibinfo {author} {\bibfnamefont {A.}~\bibnamefont
  {Glicenstein}}, \bibinfo {author} {\bibfnamefont {G.}~\bibnamefont
  {Ferioli}}, \bibinfo {author} {\bibfnamefont {N.}~\bibnamefont {\ifmmode
  \check{S}\else \v{S}\fi{}ibali\ifmmode~\acute{c}\else \'{c}\fi{}}}, \bibinfo
  {author} {\bibfnamefont {L.}~\bibnamefont {Brossard}}, \bibinfo {author}
  {\bibfnamefont {I.}~\bibnamefont {Ferrier-Barbut}},\ and\ \bibinfo {author}
  {\bibfnamefont {A.}~\bibnamefont {Browaeys}},\ }\bibfield  {title} {\bibinfo
  {title} {Collective shift in resonant light scattering by a one-dimensional
  atomic chain},\ }\href {https://doi.org/10.1103/PhysRevLett.124.253602}
  {\bibfield  {journal} {\bibinfo  {journal} {Phys. Rev. Lett.}\ }\textbf
  {\bibinfo {volume} {124}},\ \bibinfo {pages} {253602} (\bibinfo {year}
  {2020})}\BibitemShut {NoStop}%
\bibitem [{Note1()}]{Note1}%
  \BibitemOpen
  \bibinfo {note} {The largest duration of the pulse we used is $\SI
  {150}{\nano \second }$ and the typical thermal velocity is $v_{th}\simeq \SI
  {0.25}{\meter /\second }$. The thermal expansion during the driving is
  $\simeq 0.05\lambda _0$ and thus the atoms can be considered as
  frozen.}\BibitemShut {Stop}%
\bibitem [{\citenamefont {Santos}\ \emph {et~al.}(2021)\citenamefont {Santos},
  \citenamefont {Cidrim}, \citenamefont {Villas-Boas}, \citenamefont {Kaiser},\
  and\ \citenamefont {Bachelard}}]{Santos2021}%
  \BibitemOpen
  \bibfield  {author} {\bibinfo {author} {\bibfnamefont {A.~C.}\ \bibnamefont
  {Santos}}, \bibinfo {author} {\bibfnamefont {A.}~\bibnamefont {Cidrim}},
  \bibinfo {author} {\bibfnamefont {C.~J.}\ \bibnamefont {Villas-Boas}},
  \bibinfo {author} {\bibfnamefont {R.}~\bibnamefont {Kaiser}},\ and\ \bibinfo
  {author} {\bibfnamefont {R.}~\bibnamefont {Bachelard}},\ }\href@noop {}
  {\bibinfo {title} {Generating long-lived entangled states with free-space
  collective spontaneous emission}} (\bibinfo {year} {2021}),\ \Eprint
  {https://arxiv.org/abs/2110.15033} {arXiv:2110.15033 [quant-ph]} \BibitemShut
  {NoStop}%
\bibitem [{\citenamefont {Masson}\ \emph {et~al.}(2020)\citenamefont {Masson},
  \citenamefont {Ferrier-Barbut}, \citenamefont {Orozco}, \citenamefont
  {Browaeys},\ and\ \citenamefont {Asenjo-Garcia}}]{masson2020}%
  \BibitemOpen
  \bibfield  {author} {\bibinfo {author} {\bibfnamefont {S.~J.}\ \bibnamefont
  {Masson}}, \bibinfo {author} {\bibfnamefont {I.}~\bibnamefont
  {Ferrier-Barbut}}, \bibinfo {author} {\bibfnamefont {L.~A.}\ \bibnamefont
  {Orozco}}, \bibinfo {author} {\bibfnamefont {A.}~\bibnamefont {Browaeys}},\
  and\ \bibinfo {author} {\bibfnamefont {A.}~\bibnamefont {Asenjo-Garcia}},\
  }\bibfield  {title} {\bibinfo {title} {Many-body signatures of collective
  decay in atomic chains},\ }\href
  {https://doi.org/10.1103/PhysRevLett.125.263601} {\bibfield  {journal}
  {\bibinfo  {journal} {Phys. Rev. Lett.}\ }\textbf {\bibinfo {volume} {125}},\
  \bibinfo {pages} {263601} (\bibinfo {year} {2020})}\BibitemShut {NoStop}%
\bibitem [{\citenamefont {Asenjo-Garcia}\ \emph {et~al.}(2017)\citenamefont
  {Asenjo-Garcia}, \citenamefont {Moreno-Cardoner}, \citenamefont {Albrecht},
  \citenamefont {Kimble},\ and\ \citenamefont {Chang}}]{asenjo2017exponential}%
  \BibitemOpen
  \bibfield  {author} {\bibinfo {author} {\bibfnamefont {A.}~\bibnamefont
  {Asenjo-Garcia}}, \bibinfo {author} {\bibfnamefont {M.}~\bibnamefont
  {Moreno-Cardoner}}, \bibinfo {author} {\bibfnamefont {A.}~\bibnamefont
  {Albrecht}}, \bibinfo {author} {\bibfnamefont {H.~J.}\ \bibnamefont
  {Kimble}},\ and\ \bibinfo {author} {\bibfnamefont {D.~E.}\ \bibnamefont
  {Chang}},\ }\bibfield  {title} {\bibinfo {title} {Exponential improvement in
  photon storage fidelities using subradiance and ``selective radiance'' in
  atomic arrays},\ }\href {https://doi.org/10.1103/PhysRevX.7.031024}
  {\bibfield  {journal} {\bibinfo  {journal} {Phys. Rev. X}\ }\textbf {\bibinfo
  {volume} {7}},\ \bibinfo {pages} {031024} (\bibinfo {year}
  {2017})}\BibitemShut {NoStop}%
\bibitem [{\citenamefont {Olmos}\ \emph {et~al.}(2014)\citenamefont {Olmos},
  \citenamefont {Yu},\ and\ \citenamefont {Lesanovsky}}]{olmos2014steady}%
  \BibitemOpen
  \bibfield  {author} {\bibinfo {author} {\bibfnamefont {B.}~\bibnamefont
  {Olmos}}, \bibinfo {author} {\bibfnamefont {D.}~\bibnamefont {Yu}},\ and\
  \bibinfo {author} {\bibfnamefont {I.}~\bibnamefont {Lesanovsky}},\ }\bibfield
   {title} {\bibinfo {title} {Steady-state properties of a driven atomic
  ensemble with nonlocal dissipation},\ }\href
  {https://doi.org/10.1103/PhysRevA.89.023616} {\bibfield  {journal} {\bibinfo
  {journal} {Phys. Rev. A}\ }\textbf {\bibinfo {volume} {89}},\ \bibinfo
  {pages} {023616} (\bibinfo {year} {2014})}\BibitemShut {NoStop}%
\bibitem [{\citenamefont {Parmee}\ and\ \citenamefont
  {Cooper}(2018)}]{parmee2018phases}%
  \BibitemOpen
  \bibfield  {author} {\bibinfo {author} {\bibfnamefont {C.~D.}\ \bibnamefont
  {Parmee}}\ and\ \bibinfo {author} {\bibfnamefont {N.~R.}\ \bibnamefont
  {Cooper}},\ }\bibfield  {title} {\bibinfo {title} {Phases of driven two-level
  systems with nonlocal dissipation},\ }\href
  {https://doi.org/10.1103/PhysRevA.97.053616} {\bibfield  {journal} {\bibinfo
  {journal} {Phys. Rev. A}\ }\textbf {\bibinfo {volume} {97}},\ \bibinfo
  {pages} {053616} (\bibinfo {year} {2018})}\BibitemShut {NoStop}%
\bibitem [{\citenamefont {Parmee}\ and\ \citenamefont
  {Ruostekoski}(2020)}]{parmee2020signatures}%
  \BibitemOpen
  \bibfield  {author} {\bibinfo {author} {\bibfnamefont {C.~D.}\ \bibnamefont
  {Parmee}}\ and\ \bibinfo {author} {\bibfnamefont {J.}~\bibnamefont
  {Ruostekoski}},\ }\bibfield  {title} {\bibinfo {title} {{Signatures of
  optical phase transitions in superradiant and subradiant atomic arrays}},\
  }\href {https://doi.org/10.1038/s42005-020-00476-1} {\bibfield  {journal}
  {\bibinfo  {journal} {Communications Physics}\ }\textbf {\bibinfo {volume}
  {3}},\ \bibinfo {pages} {205} (\bibinfo {year} {2020})},\ \Eprint
  {https://arxiv.org/abs/2007.03473} {2007.03473} \BibitemShut {NoStop}%
\bibitem [{\citenamefont {Lewis-Swan}\ \emph {et~al.}(2021)\citenamefont
  {Lewis-Swan}, \citenamefont {Muleady}, \citenamefont {Barberena},
  \citenamefont {Bollinger},\ and\ \citenamefont
  {Rey}}]{lewis2021characterizing}%
  \BibitemOpen
  \bibfield  {author} {\bibinfo {author} {\bibfnamefont {R.~J.}\ \bibnamefont
  {Lewis-Swan}}, \bibinfo {author} {\bibfnamefont {S.~R.}\ \bibnamefont
  {Muleady}}, \bibinfo {author} {\bibfnamefont {D.}~\bibnamefont {Barberena}},
  \bibinfo {author} {\bibfnamefont {J.~J.}\ \bibnamefont {Bollinger}},\ and\
  \bibinfo {author} {\bibfnamefont {A.~M.}\ \bibnamefont {Rey}},\ }\bibfield
  {title} {\bibinfo {title} {Characterizing the dynamical phase diagram of the
  dicke model via classical and quantum probes},\ }\href
  {https://doi.org/10.1103/PhysRevResearch.3.L022020} {\bibfield  {journal}
  {\bibinfo  {journal} {Phys. Rev. Research}\ }\textbf {\bibinfo {volume}
  {3}},\ \bibinfo {pages} {L022020} (\bibinfo {year} {2021})}\BibitemShut
  {NoStop}%
\end{thebibliography}%

\end{document}